%% file: HfKID.tex
\documentclass[aip,amsmath,amssymb,reprint]{revtex4-1}
\usepackage{graphicx}
\usepackage{dcolumn}
\usepackage{bm}

\usepackage[utf8]{inputenc}
\usepackage[T1]{fontenc}
\usepackage{mathptmx}
\usepackage{etoolbox}

\usepackage{siunitx}
\usepackage{verbatim}

\draft 

\begin{document}


\title{Low Tc Hafnium Kinetic Inductance Device with High Internal Quality Factor} 

\author{Xinran Li}
\email[Email: ]{xinranli@lbl.gov}
\affiliation{Lawrence Berkeley National Laboritory, Berkeley, CA, USA}
\author{Aritoki Suzuki}
\affiliation{Lawrence Berkeley National Laboritory, Berkeley, CA, USA}
\author{Maurice Garcia-Sciveres}
\affiliation{Lawrence Berkeley National Laboritory, Berkeley, CA, USA}

\date{\today}
\begin{abstract}
Kinetic inductance devices (KIDs) are superconducting resonators with high kinetic inductance sensitive to external energy perturbations. KIDs made with superconductors having $T_c$ far below one Kelvin are of particular interest for sensing minuscule signals, such as light dark matter detection and millimeter wave telescopes for astronomy and cosmology. In this work, we report the promising performance of KIDs fabricated with Hafnium from heated sputter depositions. The KIDs have $T_c$ lower than \SI{249(2)}{\milli\kelvin}, the internal quality factor ($Q_i$) of the resonators exceeds $10^5$ and the temperature dependence of the resonances can be described by the generalized Mattis-Bardeen model with a disorder parameter $\Gamma=4\times10^{-3}\Delta_0$. Post-fabrication annealing at temperatures above the deposition temperature can further reduce the $T_c$ and $\Gamma$ without reducing $Q_i$, leading to improvements in energy resolution. 
\end{abstract}







\maketitle 


\input{introduction}

\label{sec:setup}
\input{setup}

\label{sec:linear}
\input{linearResults}

\label{sec:conclusion}
\input{conclusion}

\section{Supplementary Material}
Details of the RF circuit setup, estimation of the nonlinear driving power of the KIDs, the fitting function of $S_{21}$, and the generalization of the M-B model are included in the supplementary material. 

\begin{acknowledgments}
We thank the Nanofabrication facility at the Molecular Foundry in Lawrence Berkeley National Laboratory for sample annealing, the SEEQC, inc. for microfabrication, and the STAR Cryoelectronics, inc. for hafnium film deposition.This work is supported by the U.S. DOE, Office of High Energy Physics and Quantum Information Science Enabled Discovery (QuantISED) for High Energy Physics, Office of Science, Office of Basic Energy Sciences, under Contract DE-AC02-05CH11231. 
\end{acknowledgments}
\bibliography{HfKID}

\end{document}


\title{Low Tc Hafnium Kinetic Inductance Device with High Internal Quality Factor - Supplementary Material}

\author{Xinran Li}
\email[Email: ]{xinranli@lbl.gov}
\affiliation{Lawrence Berkeley National Laboritory, Berkeley, CA, USA}
\author{Aritoki Suzuki}
\affiliation{Lawrence Berkeley National Laboritory, Berkeley, CA, USA}
\author{Maurice Garcia-Sciveres}
\affiliation{Lawrence Berkeley National Laboritory, Berkeley, CA, USA}

\date{\today}

\maketitle

\section{RF circuit.}
The 12 KIDs of the two bands were grouped on two chips and readout by an LNF-LNA0.2-3B and an LNF-LNA4-8F HEMT amplifier separately. Details of the readout circuit are shown in Fig.~1.A of the main article. Extra attention was exercised to reduce parasitic RF power loading on the low-Tc KIDs. A total of \SI{73}{dB} attenuation was added to the input line to minimize thermal photons emitted at higher temperatures from reaching the KIDs. Eccosorb filters (Quantum Microwave QMC-CRYOIRF-003MF-S) were installed on both sides of the test box to filter photons at \SI{10}{\giga\hertz} and above. The high-frequency band had a two-junction circulator (QuinStar QCI-G0400802) terminated at \si{\milli\kelvin} to reduce back-propagating thermal noise from the HEMT. No suitable circulator was found for the low-frequency band. Instead, extra lumped element low-pass filters with \SI{3.8}{\giga\hertz} cut-off frequencies were added. Finally, a \SI{23}{\decibel} gain room temperature amplifier is used to increase the signal to VNA readout noise ratio and reduce the data taking average time.

We attribute the $10\sim\SI{15}{\decibel}$ additional attenuation measured on the VNA at off-resonance frequencies to cable loss and assume half of it occurs before the chip.

\section{KID nonlinear driving power estimation.} 
The critical field is reached when the stored energy in the resonator is comparable to the condensate energy \cite{swenson2013operation, tinkham2004introduction}
\begin{equation}
    \frac{P_\mathrm{feed} Q_r}{2\pi f_r}\sim2N_0\Delta_0^2V_L
\end{equation}
where $Q_r$ and $f_r$ are the total quality factor and resonance frequency, respectively, $N_0$ is the single-spin density of electron states at the Fermi energy, $N_0=\SI{3.6e10}{\per\cubic\micro\meter\per\electronvolt}$ for Hf\cite{jepsen1975electronic}, $\Delta_0$ is the superconducting gap at zero temperature, and $V_L$ is the inductor volume. The resulting power value is roughly \SI{-50}{dBm}. On the other hand, heating the quasiparticles to \SI{40}{\milli\kelvin} results in significant changes in $f_r$ and $Q_r$ according to the top panels in Fig.~2 in the main article. Dissipation through quasiparticle-phonon coupling is significantly suppressed at low temperatures as \cite{wellstood1994hot, de2010readout}
\begin{equation}
    P_\mathrm{el-ph} = V_L\Sigma(T_\mathrm{el}^5-T_\mathrm{ph}^5)
\end{equation}
where $T_\mathrm{el}$ and $T_\mathrm{ph}$ are the electron (quasiparticle) temperature and phonon temperature, respectively, and $\Sigma = 10^{8}\sim10^{9}\si{\watt\meter^{-3}\kelvin^{-5}}$ is the electron-phonon coupling constant.\footnote{Estimated with the value of aluminum \cite{de2010readout}, assuming it is on the same order of magnitude in pure metals.} 
The resulting power is \SI{-110}{dBm} to \SI{-100}{dBm}, matches well with the measurement.

\section{Fitting function for $S_{21}$ to extract $Q_i$ and $f_r$}
We use the following Eq.~\ref{eq:S21}
\begin{equation}
    S_{21}(f) = |a| e^{i\theta_a} e^{-i2\pi\tau f}\left( 1 - \frac{e^{i\phi}Q_r/Q_c}{\cos{\phi}(1+2iQ_r(1-f/f_r))}  \right)
    \label{eq:S21}
\end{equation}
where $|a|$ and $\theta_a$ are the overall gain and phase delay of the transmission line, including all the attenuation and amplification. Here, $\tau$ is the propagation time in the transmission line, which will introduce a phase shift proportional to the frequency $f$. The formula in the brackets is the simple resonant response when $\phi=0$. The additional phase $\phi$ accounts for the skew of the resonance, or equivalently, a rotation around $(0,0.5)$ on the I-Q complex plane due to imperfections in the transmission line. Finally, $Q_c$ is the coupling quality factor and $Q_r=(1/Q_c+1/Q_i)^{-1}$ is the total quality factor. Formulas of the same format but slightly different notations were also suggested by previous works\cite{probst2015efficient}. Here, the $\cos{\phi}$ term on the denominator removes the $\phi$ dependence in the off-resonance transmission amplitude $|a|$ and makes the fitting more robust.

\section{Generalization of the Marttis-Bardeen model} 
The temperature dependence of $Q_i$ and $f_r$ follows the Mattis-Bardeen (M-B) theory on the complex conductivity of superconducting films \cite{mattis1958theory}. Specifically, we find that for Hf, a gap-broadening factor $\Gamma$ can be introduced to better fit the data \cite{nam1967theory, vzemlivcka2015finite, zobrist2019design}. For clarity, we reproduce the equations we adopted below, which can also be found in \cite{vzemlivcka2015finite}. The conductivities are 
\begin{eqnarray}
    \frac{\sigma_1}{\sigma_n} &=& \frac{1}{\hbar \omega} \int_{-\infty}^{\infty} \left[ f(E) - f(E + \hbar \omega) \right] \\
    &&\!\!\!\!\!\!\!\!\!\!\!\!
    \left[ \Re(n(E)) \Re(n(E + \hbar \omega)) + \Re(p(E)) \Re(p(E + \hbar \omega)) \right] dE \nonumber \\
    \frac{\sigma_2}{\sigma_n} &= &\frac{1}{\hbar \omega} \int_{-\infty}^{\infty} \left[ 1 - 2 f(E + \hbar \omega) \right] \\
    &&\!\!\!\!\!\!\!\!\!\!\!\!
    \left[ \Im(n(E)) \Re(n(E + \hbar \omega)) 
    + \Im(p(E)) \Re(p(E + \hbar \omega)) \right] dE \nonumber 
\end{eqnarray}
where $f(E,T) = 1/(1+e^{E/k_\mathrm{B}T})$ is the Fermi distribution, $\Re$ means taking the real part, $\Im$ means taking the imaginary part, and
\begin{eqnarray}
    n(E) &\equiv& \text{sgn}(E) \frac{E + i\Gamma}{\sqrt{(E + i\Gamma)^2 - \Delta^2}} \\
    p(E) &\equiv& \text{sgn}(E) \frac{\Delta}{\sqrt{(E + i\Gamma)^2 - \Delta^2}}
\end{eqnarray}
$\text{sgn}(E)$ is $-1, 0, 1$ for $E<0, E=0, E>0$, respectively, and the square roots are taken with real part $\geq0$. The superconducting gap at finite temperature, $\Delta$, is calculated by solving 
\begin{equation}
    \frac{1}{N_0V} = \int_0^{\hbar\omega_c} \frac{\tanh(\sqrt{E^2+\Delta^2}/2k_\mathrm{B}T)}{\sqrt{E^2+\Delta^2}} dE 
\end{equation}
numerically, where $\omega_c$ is a cutoff frequency that is sufficiently large to avoid divergence of the integral. Here, we choose $\hbar\omega_c=100\Delta_0$ and $\Delta_0=1.74k_\mathrm{B}T_c$ is the zero temperature superconducting gap. The total number of states $N_0V$ is estimated by setting $\Delta=\Delta_0$ and $T\to0$.

Finally, the quasiparticle density, $n_\mathrm{qp}$, can be written as 
\begin{equation}
    n_\mathrm{qp} = 4 N_0\int_0^\infty f(E)\Re(n(E)) dE
\end{equation}

\bibliography{supplementary}

%% file: introduction.tex
Kinetic inductance devices (KIDs) are superconducting resonators microfabricated from thin films with high kinetic inductance. KIDs have broad applications in astronomy \cite{day2003broadband, zobrist2019design}, cosmology \cite{dibert2022development}, and particle physics \cite{moore2012position, temples2024performance, cruciani2022bullkid}. They can be designed as quantum sensors to detect single photons \cite{guo2017counting, day202425} or multiplexing readout for detector arrays \cite{noroozian2013high, sypkens2024frequency}. 

Recently, using KIDs as dark matter direct detectors has been revitalized \cite{temples2024performance, cruciani2022bullkid} as the field shifts the focus to light dark matter (DM) candidates with mass below \SI{1}{\giga\electronvolt\per\square c}, which requires low energy threshold technologies beyond conventional ionizing or scintillating detectors \cite{kahn2022searches, hertel2019direct, fink2020characterizing}. KIDs are sensitive to $O(1)\sim O(100)\si{\milli\electronvolt}$ athermal phonons produced by energy deposited in the crystal substrates they are built on, which makes them sensitive to sub-GeV DM interactions \cite{knapen2022python}. 
Currently, the most sensitive athermal phonon detectors are built with transition edge sensors (TESs) \cite{anthony2024low}, and they have been implemented in multiple light DM experiments \cite{angloher2024first, albakry2025light, anthony2024demonstration}. Since TESs require superconducting quantum interference device (SQUID) amplifiers as the low-noise, low-impedance readout, scaling in detector mass and reading out large TES arrays is challenging.
In comparison, KIDs operate in radio frequency (RF), which is naturally multiplexible. They have the potential to achieve sub-eV energy thresholds and the scalability to thousands of channels for the next-generation kilogram detectors. 

Among the various materials that have been explored \cite{mazin2022superconducting}, materials with low critical temperatures ($T_c$) are preferred for sensing, as more quasiparticles will be produced per unit of energy. A high internal quality factor ($Q_i$) further magnifies the small change in resonance frequency. Previously, Aluminum (Al) KIDs have been widely studied with $T_c=\SI{1.2}{\kelvin}$ and consistently achieved high $Q_i$ around $10^6$. Recent works \cite{day202425} demonstrated the possibility of detecting single THz photons with Al KIDs. The responsivity is significantly boosted by reducing the inductor volume to \si{\cubic\nano\meter} scale. However, this approach requires attaching an antenna to the inductor to increase the collection area and focusing photons onto the antenna, which does not trivially apply to an athermal phonon detector.

Similarly to an antenna for electromagnetic waves, an Al phonon absorber coupling to the KID with a quasiparticle trap\cite{booth1987quasiparticle} can collect phonons, but it requires even lower $T_c$ KIDs. Quasiparticle trapping significantly improves phonon collection efficiency while maintaining a small inductor volume and high responsivity. This technique has been successfully demonstrated in TESs\cite{irwin1995quasiparticle, fink2021performance}, with Al as the phonon absorber and Tungsten as TESs. It also has been demonstrated in X-ray KID detectors, with Tantalum (Ta) as the photon absorber and Al as KIDs \cite{mazin2006position}. Al is perfect for the high-$T_c$ phonon absorber in a quasiparticle trap, as it has a long quasiparticle lifetime\cite{kaplan1976quasiparticle} and a matched acoustic impedance to common substrates such as Si. We are therefore searching for KID materials with $T_c$ much less \cite{kaplan1976quasiparticle} than \SI{1.2}{\kelvin}, high $Q_i$, high kinetic inductance, and fabrication compatibility for DM KID detector development. 

Low $T_c$ materials such as $\beta$-Ta, Osmium, Titanium, Hafnium (Hf), and Iridium have been explored \cite{mazin2022superconducting}, among which we consider Hf the most promising candidate due to its suitable $T_c$, high kinetic inductance, and high $Q_i$. Previous works on Hf KIDs\cite{coiffard2020characterization, zobrist2019design} have demonstrated large pixel arrays that can resolve \SI{1}{\electronvolt} IR photons. 

We have obtained promising results with KIDs fabricated from Hf using a novel heated sputter deposition process on silicon substrates \cite{rotermund2024development}. A $T_c$ of \SI{250}{\milli\kelvin} has been achieved with a surface inductance of $\SI{4.8}{\pico\henry}/\square$. The $Q_i$ of the resonators is higher than $10^5$, sufficient for use as phonon sensors. Notably, the $T_c$ can be tuned precisely post-fabrication by annealing. Adding the quasiparticle trapping technique to such Hf KIDs would lead to sub-eV threshold phonon-sensing detectors.


%% file: setup.tex
The Hf films in this study are \SI{250}{\nano\meter} thick, deposited with plasma sputtering on Si substrates heated to \SI{200}{\celsius}. We chose the thickness as it is an established fabrication process from previous work on Hf-TESs development for cosmic microwave background detectors\cite{rotermund2024development}; we will reduce the thickness in the future to achieve higher surface kinetic inductance. The $T_c$ of the film lowers linearly as the substrate temperature increases during deposition. Above \SI{550}{\celsius}, the film $T_c$ approaches bulk Hf $T_c$ of \SI{128}{\milli\kelvin} \cite{kraft1998use}. 
We chose a \SI{200}{\celsius} deposition temperature for $T_c$ at \SI{250}{\milli\kelvin}, which allowed us to reliably operate KIDs fabricated from the films in a dilution refrigerator (DR). A chlorine-based plasma etching is used to pattern the film into KIDs.

\begin{figure}
    \centering
    \includegraphics[width=\linewidth]{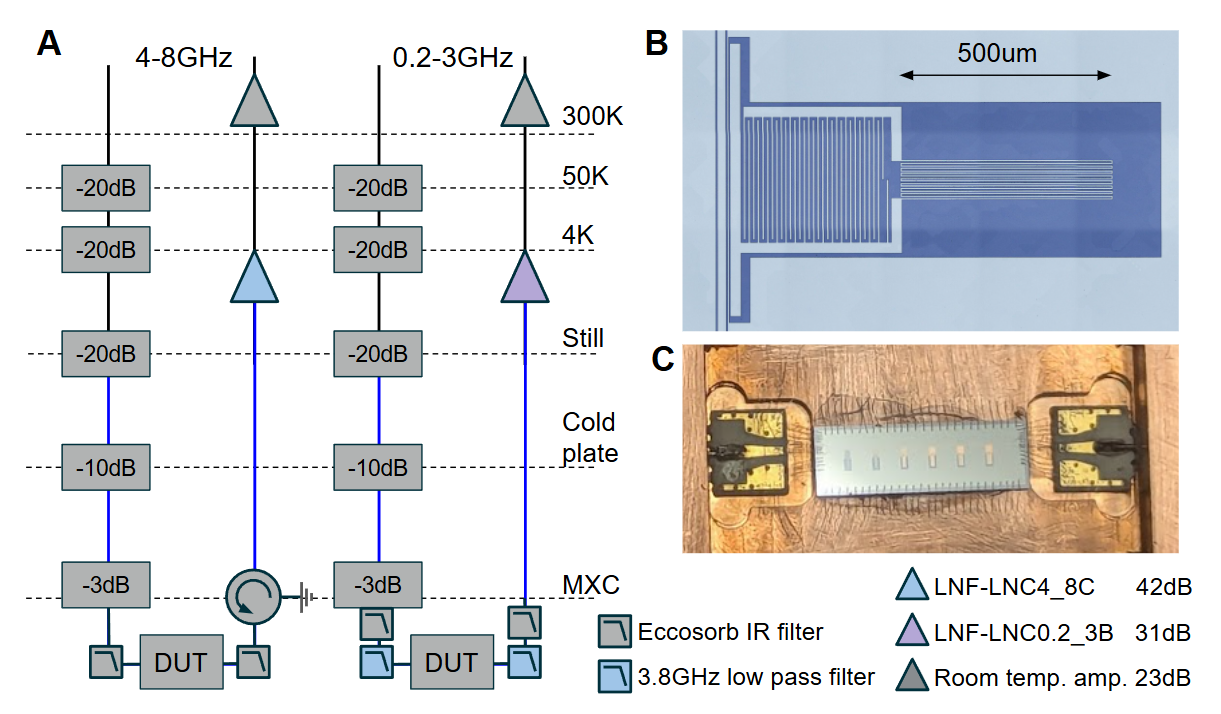}
    \caption{{\bf A}) Cold electronics of the two frequency bands. DUT refers to the KID housed in the copper box. {\bf B}) Example microscope picture the KID. Light blue is Hf and dark blue is exposed silicon. {\bf C}) Picture of the KID chip mounted in the copper box with rubber cement.}
    \label{fig:setup}
\end{figure}

At such low $T_c$, the pair-breaking energy is $2\times 1.76 k_B T_c=\SI{76}{\micro\electronvolt}$, corresponding to \SI{18.4}{\giga\hertz} photons. The KID frequency should be as low as possible to minimize parasitic quasiparticle excitations by the readout power \cite{goldie2012non}. Here, we designed 12 lumped element KIDs with resonances from \SI{1.8}{\giga\hertz} up to \SI{6}{\giga\hertz} to test the quasiparticle dynamics under different readout frequencies. They share the same interdigitated capacitor design and vary in the inductor dimensions. They are distributed across two chips, the six resonators in the lower (higher) frequency band have \SI{8}{\milli\meter} (\SI{2}{\milli\meter}) long meandering inductors with width ranging from \SI{3}{\micro\meter} to \SI{12}{\micro\meter}. Fig.~\ref{fig:setup}.B shows the \SI{1.8}{\giga\hertz} KID design as an example. 

The KIDs were cooled by a DR and readout with two high electron mobility transistor (HEMT) amplifiers. Extra attention was exercised to reduce parasitic RF power loading on the low-Tc KIDs. Details of the RF circuit is shown in Fig.~\ref{fig:setup}.A, and further described in the supplementary material.
The chips were mounted in an IR light-tight copper box using rubber cement. Wire bonds were added from the ground plane to the copper box to ensure good grounding, as shown in Fig.~\ref{fig:setup}.C.

\begin{figure}
    \centering
    \includegraphics[width=\linewidth]{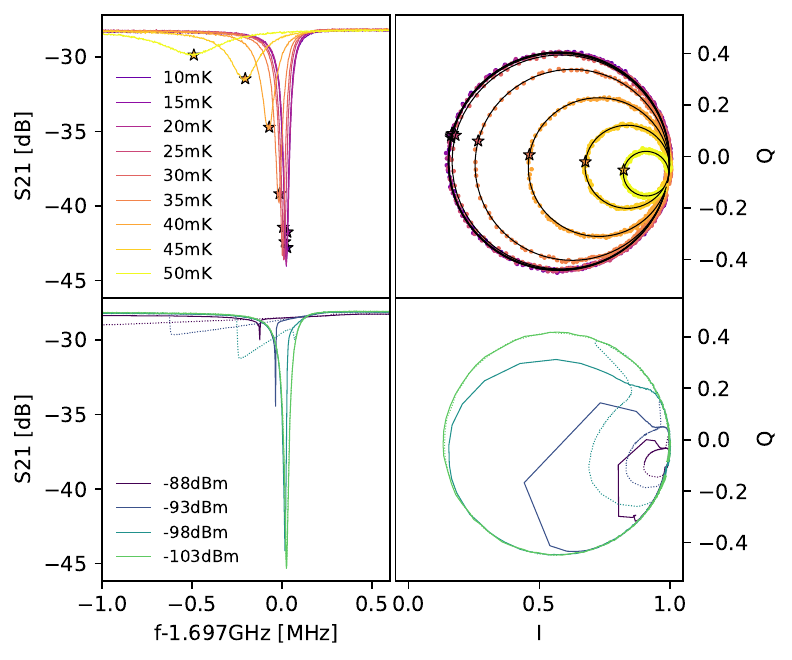}
    \caption{Example S21 (left panels) and I-Q (right panels) scan for the KID in Fig.~\ref{fig:setup}B. Top panels show the scan with \SI{-113}{dBm} readout power at various bath temperatures. Black curves are the best model fits; the star marks the $f_r$ point. The bottom panels show the scan with various readout power at \SI{20}{\milli\kelvin}. Solid lines are ascending frequency scans, dashed lines are descending frequency scans. Differences in the two scanning directions due to quasiparticle heating are observed with $P_\mathrm{feed}>\SI{103}{dBm}$.}
    \label{fig:S21andIQ}
\end{figure}

A vector network analyzer was used to measure the I-Q response of signal transmission, $S_{21}$. 
An example scan of the lowest frequency KID is shown in Fig.~\ref{fig:S21andIQ}. The top panels show the dependence on temperature while the readout power is relatively low, \SI{-113}{dBm} on the feedline. This power, $P_\mathrm{feed}$, is estimated with cable loss corrections. Unless specified otherwise, all absolute power values mentioned hereafter refer to the power on the feedline, and they are associated with $\pm\SI{2}{\decibel}$ uncertainty. After each scan, we fit the resonant circle in the I-Q plane to extract $Q_i$ and $f_r$; see supplement material for the fitting function.

The bottom panels of Fig.~\ref{fig:S21andIQ} show the scans of the same KID with higher powers. Significant nonlinearity is observed for powers above \SI{-103}{dBm}, which results in hysteresis when scanning frequencies in different directions. Two known effects can induce nonlinearity: one is a large current generating magnetic field that changes the gap energy \cite{swenson2013operation}, and the other is quasiparticle heating by the readout power \cite{de2010readout}. A rough estimation suggests that the second effect occurs first in the low-Tc Hf KID; see supplementary material for details.

%% file: linearResults.tex
We restrict the discussion below to the linear regime. The lowest frequency KID is most easily saturated, which shows hysteresis around \SI{-98}{dBm}. The following results are restricted to $P_\mathrm{feed}\leq\SI{-110}{dBm}$, so that the quasiparticles are not significantly disturbed.

The most essential parameters are the kinetic inductance of the Hf film, $L_k$, and the internal quality factor, $Q_i$. We estimated $L_k$ by comparing the measured resonant frequencies at \SI{10}{\milli\kelvin} to simulation results. We used \texttt{SONNET}\cite{SONNET2018} to simulate the RF field with $L_k$ parametrized as the surface inductance of a perfect conductor\cite{kerr1999surface}. The best agreement across all KIDs was achieved with $L_k=\SI{4.8(1)}{\pico\henry}/\square$. For comparison, we can estimate $L_k$ from the normal resistance, assuming the thin film limit \cite{zmuidzinas2012superconducting}.
\begin{equation}
    L_k = \mu_0 \lambda_\mathrm{thin} \approx \mu_0 \frac{\SI{105}{\nano\meter}^2}{t} \frac{R_n t}{\SI{1}{\mu\Omega\centi\meter}}\frac{\SI{1}{\kelvin}}{T_c}
    \label{eq:Lk}
\end{equation}
where $\mu_0=4\pi\times10^{-7}\si{\henry/\meter}$ is the vacuum permeability, $t=\SI{250}{\nano\meter}$ is the film thickness, and $R_n$ is the normal sheet resistance. We measured $T_c=\SI{249(2)}{\milli\kelvin}$ and $R_n=\SI{0.86(3)}{\Omega}/\square$ at \SI{1}{\kelvin} with an AC resistance bridge. The resulting $L_k$ is $\SI{4.8(1)}{\pico\henry}/\square$, agrees well with the measurement from resonances. 

The $T_c$ of Hf can be tuned post-fabrication with annealing at temperatures above the original deposition temperature. We annealed one set of chips at \SI{300}{\degreeCelsius} in an Argon environment for $30$ minutes. The $T_c$ of the annealed chip dropped to \SI{233(2)}{\milli\kelvin} while the normal resistance decreased to $\SI{0.83(3)}{\Omega}/\square$. Eq.~\ref{eq:Lk} suggests $L_k$ increases to $\SI{4.9(1)}{\pico\henry}/\square$ and the measured value is $\SI{5.3(1)}{\pico\henry}/\square$. 

\begin{figure}
    \centering
    \includegraphics[width=\linewidth]{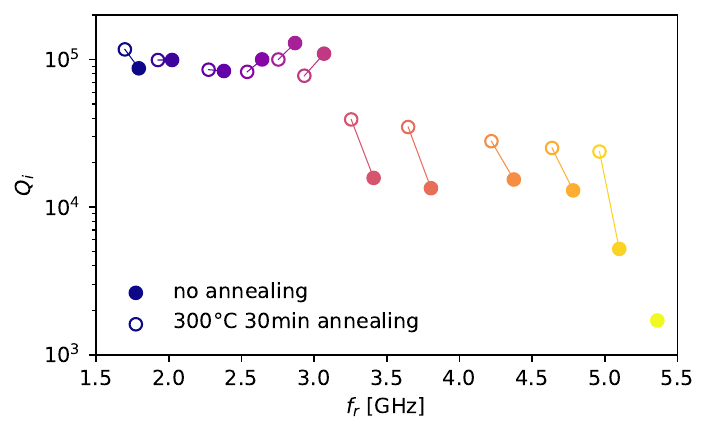}
    \caption{$Q_i$ at \SI{10}{\milli\kelvin} and \SI{-118}{dBm} readout power of KIDs with (open circle) and without (closed circle) annealing. KIDs of the same design are connected by thin lines to guide the eye. The highest frequency resonator on the annealed chip had a defect, and no resonance was observed.}
    \label{fig:Qivsfr}
\end{figure}

The $Q_i$ values at \SI{10}{\milli\kelvin} are summarized in Fig.~\ref{fig:Qivsfr}. KIDs in the low-frequency band have $Q_i$ close to or exceeding $10^5$, which will provide excellent energy resolution as phonon sensors. No significant reduction of $Q_i$ was observed with annealing. Rather, annealed KIDs in the high-frequency band have higher $Q_i$, which is most likely caused by systematics in the measurement environment, since the no-annealing data was taken five days after the DR reached \SI{1}{\kelvin}, while the annealed data was taken 20 days after. The circulator, which only presented in the high-frequency band, might not have reached thermal equilibrium with the DR in five days, allowing thermal photon emission to back-propagate to the KID and reduce the $Q_i$. 

The dependence of the complex conductivity on frequency and temperature, $\sigma(\omega,T) = \sigma_1(\omega,T) - i\sigma_2(\omega,T)$, defines the responsivity of KIDs. J. Gao et. al. demonstrated the equivalence of thermally and external energy excited quasiparticles, as shown by Fig. 1 in the reference \cite{gao2008equivalence}
\begin{equation}
    \frac{d\sigma_{1,2}}{dE} = \frac{dn_\mathrm{qp}}{dE} \frac{d\sigma_{1,2}}{dn_\mathrm{qp}} = \frac{dn_\mathrm{qp}}{dE} \frac{\partial\sigma_{1,2}(T)/\partial T}{\partial n_\mathrm{qp}(T)/\partial T}
    \label{eq:dsigmadE}
\end{equation}
where $E$ is the external signal energy deposition in the quasiparticle system, and $n_\mathrm{qp}$ is the quasiparticle density. 

We characterize $\sigma_{1,2}(\omega,T)$ by measuring the temperature dependence of the resonant frequencies and quality factors. Following N. Zobrist, {\it et al.}\cite{zobrist2019design}, we define the normalized fractional change of frequency, $2\delta f_r/f_{r0}/\alpha\gamma$, and the quality factor, $\delta(1/Q_i)/\alpha\gamma$, where $\delta$ means the variance relative to the zero temperature value, $\alpha$ is the kinetic inductance fraction, which can be calculated from simulation, and $\gamma=1$ in the thin film approximation.
The fractional changes directly relate to $\delta\sigma_{1,2}$ as 
\begin{eqnarray}
    2\delta f_r/f_{r0}/\alpha\gamma &=& (\sigma_2(T)-\sigma_2(0))/\sigma_2(0) \\
    \label{eq:dfr}
    \delta(1/Q_i)/\alpha\gamma &=& \sigma_1(T)/\sigma_2(T) - \sigma_1(0)/\sigma_2(0)
    \label{eq:d1overQ}
\end{eqnarray}

The Mattis-Bardeen (M-B) model \cite{mattis1958theory} give the theoretical form of $\sigma_{1,2}(T)$ and $n_\mathrm{qp}(T)$ in ideal superconductors. Here, we follow Nam's generalization to parametrize the state broadening due to disorders by the parameter $\Gamma$ \cite{vzemlivcka2015finite,nam1967theory}. See supplement materials for the detailed treatment of the generalization.

\begin{figure}
    \centering
    \includegraphics[width=\linewidth]{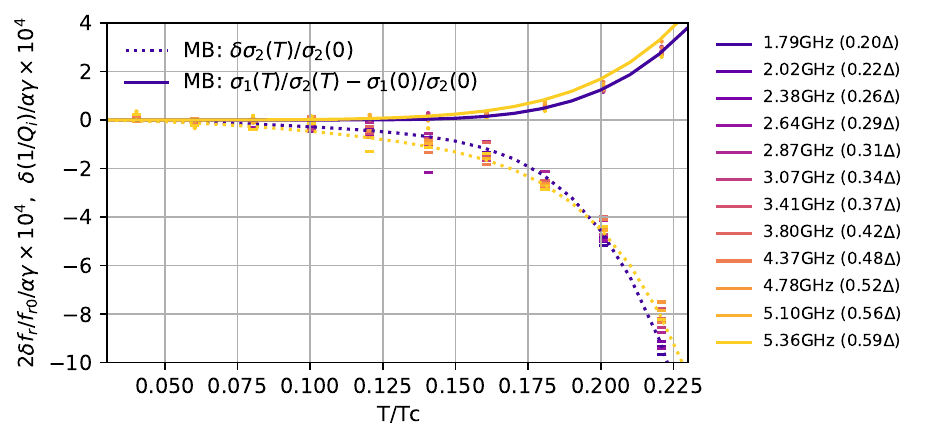}
    \includegraphics[width=\linewidth]{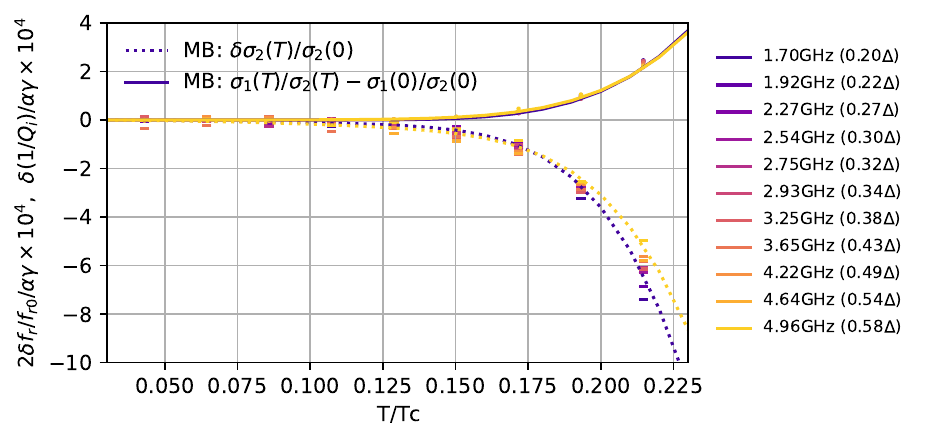}
    \caption{Normalized fractional change of resonant frequencies and internal quality factors vs temperature. Short bars and dots are the fitting results of $f_r$ and $Q_i$, respectively, for each KID at different temperatures. A darker color corresponds to KID with a lower resonant frequency. Solid and dashed curves represent the generalized M-B model of the normalized fractional change of $Q_i$ and $f_r$, respectively. {\bf Top}: Chip with no treatment. Best described with $\Gamma=4\times10^{-3}\Delta_0$. {\bf Bottom}: Chip annealed at \SI{300}{\celsius} for 30min in an Argon environment. Best described with $\Gamma=1.5\times10^{-3}\Delta_0$.}
    \label{fig:M-B}
\end{figure}

The temperature dependence of the fractional changes is shown in Fig.~\ref{fig:M-B}. KIDs without and with annealing are best fit with $\Gamma = 4\times10^{-3}\Delta_0$ and $\Gamma = 1.5\times10^{-3}\Delta_0$, respectively. The difference in $\Gamma$ is most visible around $T=0.15T_c$, where higher disorders (larger $\Gamma$) cause the frequency shift to start at lower temperatures before the $Q_i$ starts to degrade. However, this observation does not imply that a more disordered film would exhibit a stronger phase signal response. 

The model also qualitatively predicts the frequency dependence. The measurements span a wide frequency range, from $0.2\Delta/h$ to $0.6\Delta/h$, showing clear frequency dependence in $\sigma_{1,2}$. At high temperatures, high-frequency KIDs have less $\delta f_r/f_{r0}$. But at intermediate temperatures, especially for a large $\Gamma$, the trend reverses, and high-frequency KIDs have greater $\delta f_r/f_{r0}$.

Previous work \cite{zobrist2019design} on Hf KIDs used a different generalization, which also fits the data qualitatively. In this work, the best-fit suggests a superconducting gap closer to the BCS value of $1.762k_\mathrm{B}T_c$ and a smaller $\Gamma$, suggesting less disorder in the film, which is further reduced after annealing. The difference may be attributed to the different film deposition procedures.

The model predicts the energy response of the KID. 
Using Eq.~\ref{eq:dsigmadE} to~\ref{eq:d1overQ}, the perturbation of $S_{21}$ due to small energy can be written as
\begin{equation}
    \delta S_{21} = \frac{Q_r^2}{Q_c}\alpha\gamma\left[i\frac{d\sigma_2/dn_\mathrm{qp}}{\sigma_2(0)}
    + \frac{d\sigma_1/dn_\mathrm{qp}}{\sigma_2(0)}\right]\frac{dN_\mathrm{qp}}{dE}\frac{1}{V}\delta E
\end{equation}
The energy collection efficiency $\eta$ is included in the term $dN_\mathrm{qp}/dE = \eta/\Delta$, where $\eta$ includes all the efficiencies for converting the signal energy into a number of excited quasiparticles. $V$ is the sensitive volume of the KID's inductor. 

\begin{figure}
    \centering
    \includegraphics[width=\linewidth]{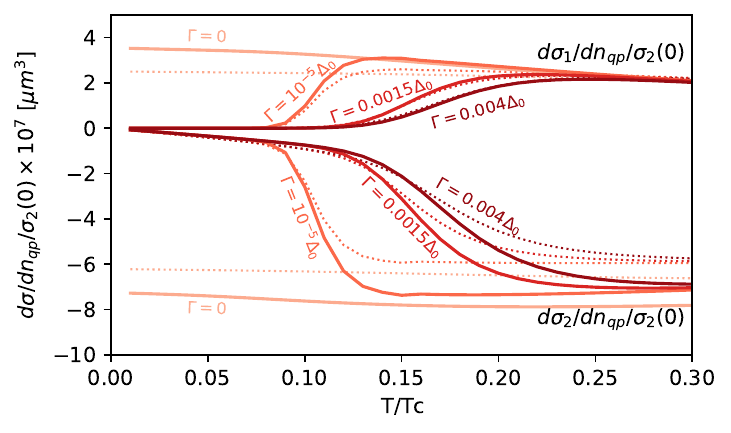}
    \caption{Signal response prediction from the generalized M-B theory. Positive and negative curves correspond to dissipation signal ($d\sigma_1/dn_\mathrm{qp}/\sigma_2(0)$) and phase signal ($d\sigma_2/dn_\mathrm{qp}/\sigma_2(0)$), respectively. Solid and dashed curves are response of KIDs at \SI{2}{\giga\hertz} and \SI{4}{\giga\hertz}, respectively. From light to dark, the disorder parameter $\Gamma$ increases. The $\Gamma=0$ curve takes the approximated analytical form from J. Gao\cite{gao2008equivalence}, as the numerical differential calculations diverge. Annealing reduces $\Gamma$ and increases signal response.}
    \label{fig:dsigmadnqp}
\end{figure}

The numerical shape of $d\sigma_{1,2}/dn_\mathrm{qp}/\sigma_2(0)$ can be derived from the model we built and tested in Fig.~\ref{fig:M-B}. Results are shown in Fig.~\ref{fig:dsigmadnqp}. Positive and negative curves show the dissipation and phase responses, respectively. In addition to the best-fit $\Gamma$ values, situations with $\Gamma=0$ and $\Gamma=10^{-5}$ are shown for reference. The solid and thin dashed curves were calculated with $f_r=\SI{2}{\giga\hertz}$ and $f_r=\SI{4}{\giga\hertz}$, respectively. Reducing the resonance frequency improves the signal response. Although $\partial\sigma/\partial T$ increases as $\Gamma$ increases, the density of states below the gap increases as well, resulting in a higher $n_\mathrm{qp}$ and higher $\partial n_\mathrm{qp}/\partial T$ that do not exponentially decay as $T$ approaches zero. In other words, more quasiparticle excitations are needed to fill the sub-gap states to generate the same effective temperature change compared to a $\Gamma=0$ ideal superconductor. Thus, a small $\Gamma$ is preferred to achieve a high signal response in a KID. In this work, the annealed chip presents a lower $\Gamma$, and it should achieve the highest response around $T=0.15\sim0.2T_c$. The $Q_i$ reduction becomes significant as $T$ approaches $0.2T_c$, and it will suppress $\delta S_{21}$ as $Q_r^2$. 

Note that the previous discussion assumes the density of states after broadening has the Lorentzian shape down to zero energy. However, other models with different distributions have been proposed \cite{skvortsov2013subgap}. The complex conductance measurement with resonators is not sensitive enough to distinguish between the various models. And the actual behavior of $\partial n_\mathrm{qp}/\partial T$ may differ. Generally speaking, any density of states broadening will increase $\partial n_\mathrm{qp}/\partial T$ close to zero temperature, reducing the response. 

Another implicit assumption requiring examination is the thermal quasiparticle distribution. Various experiments \cite{barends2008quasiparticle} have observed quasiparticle lifetimes diverging from the BCS theory as the superconductor being cooled, indicating a significant population of residue quasiparticles. Conventional treatment includes introducing an effective chemical potential \cite{gao2008equivalence}, and assuming they follow a thermal distribution. In addition, the readout microwave can also alter the quasiparticle distribution and break Cooper pairs through multiphoton processes, preventing the predicted noise equivalent power from being achieved \cite{goldie2012non}. With all the theoretical uncertainties, the prediction of KID sensitivity remains a challenging task. Further studies with energy calibrations will improve the understanding.

%% file: conclusion.tex
In conclusion, we measured the performance of KIDs fabricated from Hf films deposited on heated silicon substrates. The $T_c$ of a \SI{250}{\nano\meter} film deposited at \SI{200}{ \degreeCelsius} is \SI{249(2)}{\milli\kelvin}. Post-fabrication annealing reduces the $T_c$ to \SI{233(2)}{\milli\kelvin} and raises the sheet kinetic inductance from $\SI{4.8(1)}{\pico\henry}/\square$ to $\SI{5.3(1)}{\pico\henry}/\square$. 
The complex conductivity of Hf was carefully characterized at temperatures from \SI{10}{\milli\kelvin} to $0.25T_c$ from resonance measurements of KIDs operated at frequencies from $0.17\Delta/h$ to $0.6\Delta/h$. The temperature and frequency dependence can be well described by the generalized M-B theory with the state broadening parameter $\Gamma$. Notably, annealing also reduces $\Gamma$ from $4\times10^{-3}\Delta_0$ to $1.5\times10^{-3}\Delta_0$, which hints to an improvement in the energy response. 

We have demonstrated the potential of Hf fabricated by heated sputter deposition as a promising candidate for phonon sensing KIDs for light DM searches. Specifically, the low $T_c$ allows it to form quasiparticle trapping structures with Al phonon absorbers to improve phonon collection efficiency significantly while maintaining high signal response. Results from this work suggest annealing as a powerful tool to adjust the material properties and potentially improve the sensor performance. 
On the other hand, theoretical uncertainties in describing disordered superconductors persist. Further studies, such as quasiparticle lifetime measurement, energy calibration, and noise characterization will reveal more profound insights.